\documentclass[useAMS,usenatbib]{mn2e}

\usepackage{graphicx} 
\usepackage{comment} 
\newcommand{\ha}{\ensuremath{{\rm H}{\alpha}}}

\newcommand{\oii}{[O\,{\sc ii}]}
\newcommand{\oiii}{[O\,{\sc iii}]}
\newcommand{\nai}{Na\,{\sc i}}
\newcommand{\caii}{Ca\,{\sc ii}}
\usepackage{amssymb}

\begin{document}
\title{Optical spectroscopy and initial mass function of $z=0.4$ red galaxies}
\author[Baitian Tang and Guy Worthey]{Baitian Tang$^{1,2}$\thanks{E-mail:
btang@astro-udec.cl (BT)} 
and Guy Worthey$^{1}$\thanks{E-mail:
gworthey@wsu.edu (GW)}\\
$^{1}$Department of Physics and Astronomy, Washington State
University, Pullman, WA 99163-2814, USA\\
$^{2}$Departamento de Astronom\'{\i}a, Casilla 160-C, 
Universidad de Concepci\'{o}n, Concepci\'{o}n, Chile\\
}

\date{Accepted  . Received  ; in original form 2014}

\pagerange{\pageref{firstpage}--\pageref{lastpage}} \pubyear{ }

\maketitle

\label{firstpage}

\begin{abstract}
  
Spectral absorption features can be used to constrain the stellar
initial mass function (IMF) in the integrated light of
galaxies. Spectral indices used at low redshift are in the far red,
and therefore increasingly hard to detect at higher and higher
redshifts as they pass out of atmospheric transmission and CCD detector
wavelength windows. We employ IMF-sensitive indices at bluer
wavelengths. We stack spectra of red, quiescent galaxies around $z=0.4$,
from the DEEP2 Galaxy
Redshift Survey. The $z=0.4$ red
galaxies have 2 Gyr average ages so that they cannot be passively
evolving precursors of nearby galaxies. They are slightly enhanced
in C and Na, and slightly depressed in Ti. Split by luminosity, the
fainter half appears to be older, a result that should be checked with
larger samples in the future. We uncover no evidence for IMF evolution
between $z=0.4$ and now, but we highlight the importance of sample
selection, finding that an SDSS sample culled to select archetypal
elliptical galaxies at z$\sim$0 is offset toward a more bottom
heavy IMF. Other samples, including our DEEP2 sample, show an offset
toward a more spiral galaxy-like IMF. All samples confirm that the
reddest galaxies look bottom heavy compared with bluer ones. Sample
selection also influences age-color trends: red, luminous
galaxies always look old and metal-rich, but the bluer ones can be
more metal-poor, the same abundance, or more metal-rich, depending on
how they are selected.

\end{abstract}


\section{Introduction}
\label{sect:intro}

The stellar initial mass function [IMF; $\xi (m)$] is key in many
active research fields, such as early universe studies, galaxy
evolution, star cluster evolution, and star formation. The IMF
regulates the number distribution of stellar populations as a function
of mass, $dN = \xi (m)\ dm$, leading to impacts on the luminosity
function, integrated mass to light (M/L) ratio, and number of stellar
remnants.  Direct IMF derivations are limited by observational
capabilities and uncertainties concerning the stellar mass-luminosity
relation, stellar evolution, dynamical evolution, binary fraction,
counting statistics, and other factors. To make the IMF even harder to
study, there is the distinct possibility that the IMF might vary in
different galactic environments \citep{Larson1998, Larson2005,
  Marks2012, Hopkins2013, Chabrier2014}.

\citet{Salpeter1955} described the IMF with a power-law distribution,
$\xi (m) \propto m^{- \alpha}$, and adopted an IMF slope ($\alpha$) of
2.35 for solar neighbourhood stars near the mass of the
sun. Considering stars of very high and very low mass, the power law
does not hold, and the Galactic IMF is now considered to peak at a few
tenths of one solar mass \citep{Miller1979,
  Scalo1986,Kroupa2001,Chabrier2003}. In terms of functional forms to
model the IMF, oft-cited examples are the \citet{Kroupa2001} series of
power laws and the \citet{Chabrier2003} log-normal distribution for
$m<1~ M_{\odot}$\footnote{a power law function for $m\gtrsim 1~
  M_{\odot}$} as representative of the log-normal probability density
function of turbulent gas. Because both models are calibrated by
observational data, they show a similar distribution between 0.2 and
0.8 $M_{\odot}$ \citep{Bastian2010,2012arXiv1203.1221G,Offner2014}. A
``steeper'' IMF with more low mass stars is termed bottom heavy and a
``shallower'' IMF with more high mass stars is termed top heavy.

Extending local studies based on counting individual stars to external
galaxies, recent studies explore the universality of the IMF using
integrated light and also dynamical models. Gravity-sensitive
integrated spectral features such as the giant-sensitive \caii~triplet
and the dwarf-sensitive \nai \ and FeH Wing-Ford band may indicate
that systematic IMF variation exists as a function of stellar velocity
dispersion \citep{Cenarro2003, VanDokkum2010,Conroy2012b}.  This trend
was fit by \citet{Ferreras2013} and \citet{Spiniello2014a}, but the
single power law IMF slope ($\alpha$) of the former study is a factor
of two greater than the latter one. Minimizing the uncertainties
arising from the ingredients of stellar population (SP) models appears
crucial to future IMF study \citep{Spiniello2014b}.  Another popular
approach for studying unresolved stellar systems is using dynamical
models with a dark matter halo involved. The IMF is estimated by
comparing the M/L ratios of the dynamical models and the stellar
population models \citep{Auger2010, Cappellari2012, Dutton2012,
  Posacki2015}. For example, \citet{ Cappellari2012} concluded an
universal IMF is inconsistent with early-type galaxies (ETGs),
although this kinematic result is unable to distinguish between more
stellar remnants (from a top heavy IMF) and relatively more low mass
stars (from a bottom heavy IMF).

Simple chemical evolution of a time-independent, bottom-heavy IMF
suggests a small number of stellar remnants and also implies lower
metallicities for the fossil stars. In empirical rebuttal,
observations suggest super-solar metallicities in elliptical galaxies
e.g., \citealt{Trager2000b,Trager2000a,Tang2009}). Also, the number of
stellar remnants is probably not small.  \citet{Kim2009} suggested the
number of low-mass X-ray binaries in three nearby elliptical galaxies
with mass about $10^{11}$ M$_{\odot}$ is similar to that of the Milky
Way.  \citet{Peacock2014} found a constant number\footnote{Scaled by
  the amount of K-band stellar light covered} of black holes and
neutron stars among eight different mass ETGs\footnote{To connect
  black holes and other stellar remnants to the low-mass IMF posits a
  very coherent and simple functional form for the IMF. In nature, the
  high mass end of the IMF may be independent of the low mass
  IMF.}. To reconcile these facts with a bottom-heavy IMF,
\citet{Weidner2013a,Weidner2013b} simulated galaxy evolution with a
time-dependent IMF, in which the IMF slope steepens as the star
formation rate decreases gradually (Also see \citealt{Gargiulo2015}
for a semi-analytical model with similar IMF slope assumption).

IMF evolution, if it occurs in nature, may potentially be detected
through studies of distant galaxies, and to possibly detect such is
the motivation for the present paper. In the formation scenario that
giant elliptical galaxies formed very eary in cosmic history and have
been passively evolving ever since, the IMF cannot change over time,
though the IMF for elliptical galaxies can be very different from
spiral galaxies. In the hierarchical galaxy formation picture,
elliptical galaxies are the end result of merger trees that start with
irregular and spiral galaxies as basic ingredients, and the elliptical
galaxy should be the sum of its parts and have an IMF similar to a
spiral galaxy. Since that conclusion seems incorrect, it may be that
elliptical galaxies are also subject to star formation events
periodically (gas rich mergers or gas accretion) but star formation
does not linger, but is rapidly quenched. In this variant of the
hierarchical formation scenario, elliptical galaxies may develop their
own characteristic IMF, and also, that IMF may evolve over cosmic
time; early ellipticals will have a spiral-like IMF, but as time and
mass-buildup goes on and they become more idiosyncratic, the
elliptical galaxies may develop a distinct elliptical-flavoured IMF
due to the altered star formation environment.

There is some theoretical support for expecting an evolving IMF.

\subsection{IMF theories}
\label{sect:d2}

Logically, the steeper IMFs of ETGs were either imposed at an early
creation epoch or have evolved over time. Present-day ETGs are
unlikely to have suffered buildups of \textit{only} low-mass stars in
the last third of the universe. We see no evidence of such an odd star
formation mode ongoing, c.f. the case study of giant elliptical galaxy
NGC 5128 \citep{2015ApJ...802...88N}, caught in a gas accretion event,
which emits plenty of UV light, indicating that massive stars are
forming.

Historically, the simple Jeans mass model and the turbulent Jeans
mass model were influential.
The simple Jeans mass model hypothesizes that the peak mass of the IMF 
is simply a reflection of the mean Jeans mass. For example, the Larson
model \citep{Larson1998,Larson2005} showed that the Jeans mass varies either
as $T^{3/2} \rho^{-1/2}$ or as $T^{2} P^{-1/2}$, where $T$ is temperature, 
$\rho$ is density, and $P$ is pressure. In the early Universe,
 the high cosmic background temperature, low metallicity, and intense 
 radiation from young stars and core-collapse supernova 
inevitably increase the proto-stellar cloud temperature. 
But the relation between $T$ and $\rho$
(or $P$) is still needed to estimate the Jeans mass. 
\citet{Larson2005} noted the $T-\rho$ relation
drawn from the observational and theoretical results of that time:
\begin{equation}
T=4.4\,\rho_{18}^{-0.27}K,\,\,\,\,\,\, \rho<10^{-18}\,g\,cm^{-3} 
\end{equation}
\begin{equation}
T=4.4\,\rho_{18}^{+0.07}K,\,\,\,\,\,\, \rho>10^{-18}\,g\,cm^{-3}
\end{equation} 
where $\rho_{18}$ is the density in units of $10^{-18}$ g cm$^{-3}$.
Therefore, Larson suggested the early universe favors a top-heavy IMF,
where a low density $T-\rho$ relation was assumed.

The turbulent Jeans mass model links the characteristic
mass of stars to the galactic-scale processes 
responsible for setting the characteristic temperatures and
linewidth-size relations of molecular clouds \citep{Krumholz2014}.
The most representative models are given by \citet{Chabrier2014} 
and \citet{Hopkins2013}. These models are widely cited, partly
due to the consistency between their model prediction and recent observations:
a bottom-heavy IMF for massive early type galaxy.

However, as pointed out by \citet{Krumholz2014}, the mass of the IMF
peak in the Hopkins model can be expressed as:
\begin{equation}
M_{peak} \approx \frac{c_{s}^{4}}{Q\,G^2\Sigma}
\end{equation} 
where $c_s$ is the sound speed, $Q\approx 1$ is the Toomre stability
parameter for the disk, and $\Sigma$ is the gas surface density.
$c_s$ and $\Sigma$ are both large for high surface density star
formation, in which elliptical galaxies are assumed to form.  The
parameters may or may not imply a bottom-heavy IMF for ETGs.

\subsection{IMFs observed at different redshifts }
\label{sect:d3}

If ETGs periodically form stars, but the episodes are rapidly
quenched, then most ETGs will be observed in the quiescent phases of
their star formation duty cycle, and yet as a group the IMF may evolve
over time. In that case, lookback studies may uncover the drift in
IMF.

There is further research. \citet{Shetty2014} derived the mass/light
(M/L) ratios of 68 field galaxies in the redshift range of 0.7$-$0.9
with both dynamical modelling and stellar population modelling. The
comparison of (M/L)$_{dyn}$ and (M/L)$_{pop}$ implies a Salpeter IMF,
which is also possessed by nearby galaxies with similar masses.
Meanwhile, \citet{2015ApJ...798L...4M} studied the TiO$_2$ indices of
a sample of 49 massive quiescent galaxies at $0.9<z<1.5$.  The
heaviest galaxies (M$_{\ast}>10^{11.0}~ M_{\odot}$) show a
bottom-heavy IMF and lighter galaxies
($10^{10.5}<$M$_{\ast}<10^{11.0}~ M_{\odot}$) do not. They also
concluded that the IMF of massive galaxies has remained unchanged for
the last $\sim$8 Gyr.

Luminosity evolution may also betray the IMF slope in that a top-heavy
galaxy fades more rapidly than a galaxy with the standard IMF, since
the present-day luminosity of ETG mainly comes from old stellar
populations ($\sim 1~ M_{\odot}$). According to \citet{Tinsley1976},
the luminosity of an old stellar population is proportional to
$t^{-1.6+0.3\alpha}$. Therefore, a shallower IMF implies a more dramatic
luminosity change over a fixed amount of time.  In that spirit,
\citet{VanDokkum2008} compared the luminosity evolution
($\Delta\log{(M/L_B)}$) to colour evolution ($\Delta (U-V)$) for
massive galaxies in clusters at $0.02<z<0.83$. The luminosity
evolution of these observed galaxies is faster than the trend
predicted by the \citet{Maraston2005} models with a standard
IMF. Thus, the IMFs in these galaxies might be top-heavy\footnote{
  \citet{2012arXiv1203.1221G} points out that the luminosity evolution at
  $0<z<1.5$ can only constrain the IMF between 1 $M_{\odot}$ and 1.4
  $M_{\odot}$.}.

More support for a shallow IMF at high redshift comes from
\citet{Dave2008}, who successfully brought the observed and predicted
$M_{\ast}$--SFR relation into broad agreement by modelling the
characteristic mass \^{M} as a function of redshift:
\^{M}$=0.5(1+z)^2~ M_{\odot}$, where $z<2$.

This rich variety of results is fascinating, yet ambiguous.

\subsection{Observational strategy}

Integrated light spectral features that are sensitive to stellar
surface gravity are often used for studies of nearby galaxy IMFs.
M-type giants and dwarfs emit most of their light and have important
diagnostic features at red wavelengths\footnote{ e.g., \caii
  $\lambda$8600, \nai $\lambda$8190, FeH Wing-Ford band $\lambda$9900}
\citep{VanDokkum2010,Conroy2012b,Smith2012}. However, at cosmic
distances, the inevitable redshift of spectral lines and the
relatively low quality of spectra obtainable beyond 1 $\mu$m increase
the difficulty of measuring these features in distant objects. It
motivates us to look for IMF-sensitive indices in a more accessible
band.  For example, the Na D, TiO$_1$, and TiO$_2$ indices from the
Lick/IDS system \citep{Worthey1994b,Trager1998} are known to be
IMF-sensitive. In addition, several optical IMF-sensitive indices
published by \citet{Spiniello2014a} and \citet{LaBarbera2013} have
given us more options in index selection (Table \ref{tab:index}).

These indices are also sensitive to population age, overall heavy
element content, and altered abundance ratios in the elements that
give rise to the spectral features themselves, such as Na, Ti, and
Ca. Fortunately, element ratios seem fairly well constrained, judging
by the good agreement on element mixture in recent papers
\citep{Johansson2012,Conroy2014,Worthey2014}.  Additionally, the
element sensitivity problem may be eased if multiple IMF-sensitive
indices with different element sensitivities are used.

Observationally, spectra from the DEEP2 redshift survey
\citep{Newman2013}, which targeted galaxies over a broad span of
redshifts in the spectral range 6500 \AA$-$9100 \AA~using Keck
Observatory, offer themselves as an appropriate data set.  The
combination of spectral observations plus new spectral indicators may
allow the measurement of the IMF over cosmic time.

This paper is organized as follows: The procedure of stacking DEEP2
spectra is illustrated in $\S$\ref{sect:deep2}. We compare the
measured indices from these composite spectra with local measurements
and two different models in $\S$\ref{sect:comp}. The implications on
IMF evolution are discussed in $\S$\ref{sect:disc}, and then a brief
summary of the results are given in $\S$\ref{sect:con}.

\section{Spectral Reduction}
\label{sect:deep2}

\subsection{Sample Selection}

The DEEP2 Galaxy Redshift Survey utilizes the DEIMOS multi-object
spectrograph \citep{Faber2003} on the Keck II telescope. Most of the
spectra cover 6500$-$9100 \AA, with spectral resolution $R = \lambda / \Delta \lambda \sim 6000$
\citep{Newman2013}. The optical IMF-sensitive indices are defined
around 4500$-$6500 \AA, listed in Table \ref{tab:index}. In
order to match observed and emitted spectral wavelength ranges for
this index set, we chose galaxies around $z=0.4$, corresponding to a
lookback time of about 4.3 Gyr.

\begin{table*}\normalsize
\begin{center}
\caption{Optical IMF-sensitive Indices
} \label{tab:index}
\begin{tabular}{lccccc}
\hline
Index & Units & Blue Pseudo-continuum & Central Feature & Red Pseudo-continuum & Source\\
\hline
bTiO &     mag  & 4742.750-4756.500 & 4758.500-4800.000 &   4827.875-4847.875 & 2  \\
aTiO & mag &      5420.000-5442.000 &  5445.000-5600.000 & 5630.000-5655.000 & 2 \\
NaD &   \AA & 5860.625-5875.625 & 5876.875-5909.375  & 5922.125-5948.125 & 1 \\
TiO1 &   mag  & 5816.625-5849.125 &  5936.625-5994.125 &   6038.625-6103.625  & 1 \\
TiO2 &     mag  & 6066.625-6141.625 &  6189.625-6272.125 &  6372.625-6415.125 & 1 \\
TiO2$_{SDSS}$ &     mag  & 6066.625-6141.625 & 6189.625-6272.125 & 6442.000-6455.000&  3\\
CaH1  &     mag  & 6342.125-6356.500&  6357.500-6401.750 &  6408.500-6429.750  & 2 \\
\hline
\end{tabular}
\end{center}
\hspace{-1 in}1 -- \citet{Worthey1994b};  2 -- \citet{Spiniello2014a}; 3 -- \citet{LaBarbera2013}

\end{table*}

\begin{figure}
\centering
\includegraphics [scale=0.4]{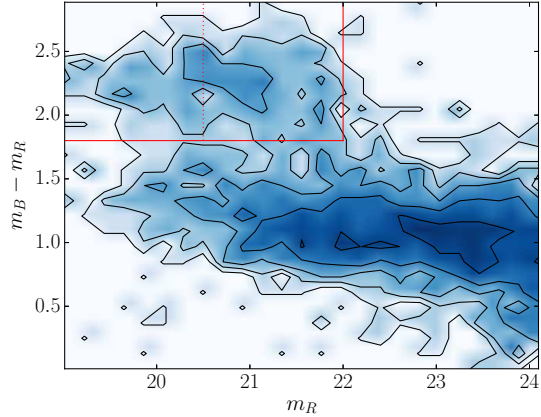} 
\caption[Galaxy colour magnitude diagram]{Galaxy colour magnitude
  diagram. Darker blue regions have greater number density. The red
  sequence is the peak in the upper left, the blue cloud is the peak
  at lower right, and the green valley is the low density gap
  between. The red galaxies are selected by $m_R <22$ mag (vertical
  red line), and $m_B-m_R>1.8$ mag (horizontal red
  line). A magnitude cut for subsampling was made at $m_R = 20.5$ mag (vertical dotted red line).}\label{fig:cmd}
\end{figure}

In detail, we selected galaxies\footnote{Confirmed by examining the
  ``CLASS'' parameter in the catalog.} with 0.3 $\leq z \leq$ 0.5 from
the Data Release 4 (DR4) redshift catalog, and made sure these
galaxies had photometric measurements by matching them with the DR4
photometric catalogs. The photometric data were taken with the CFH12K
camera on board the 3.6-meter Canada-France-Hawaii Telescope
\citep{Coil2004}. The chosen redshift range balances the number of
available IMF-sensitive indices, the sample size, and the redshift
similarity of our sample. Note that z$<$0.7 galaxies were rejected
during the observation in three out of the four DEEP2 fields (except
the Extended Groth Strip, DEEP2 field 1), and thus galaxies at
$0.3<z<0.5$ are less numerous than other higher-redshift samples
(e.g., \citealt{Schiavon2006}).  Next, to minimize the contribution of
late type galaxies (LTGs), we selected red galaxies using the galaxy
colour-magnitude diagram (CMD). We plot the ($B-R$) vs. $R$ CMD in
Figure \ref{fig:cmd}. 302 galaxies have ($B-R$) colour redder than 1.8
mag and R band magnitude brighter than $R =22$ mag. The redshifts of
these galaxies are reasonably well determined: 260 galaxies have
quality code 4 (99.5\% reliability rate), and 42 galaxies have quality
code 3 (95\% reliability rate, \citealt{Newman2013}).

According to \citet{Weiner2005}, at $z\leq1$, LTGs comprise about 25\%
of the red population. Since we lack morphology information, we cannot
exclude LTGs on the basic of isophotal structure. However, we may use
\oiii~and \ha~emission lines to eliminate galaxies with strong
emission lines \citep{Weiner2005, Schiavon2006} and therefore ongoing
star formation or active galactic nuclear activity. Based on the
wavelength coverage of the deredshifted spectra,
\oiii~EWs\footnote{Defined in \citet{Gonzalez1993}} can be measured in
282 galaxies. Among those, 7 galaxies have \oiii~EW$<-$5
\AA\footnote{\citet{Weiner2005} showed the median error in rest-frame
  EW is 6.2 \AA. \citet{Schiavon2006} suggested $-5$ \AA~for the EW
  limit, though they use \oii~due to a different rest-frame wavelength
  range.}. After \oiii selection, 275 galaxies are left in the sample
(Sample I).  Similarly, 5 out of 77 galaxies with \ha~EW measurements
have \ha~EW$<-$5 \AA, and we define thus an \ha~selected sample
(Sample II) consisting of 72 galaxies.

With a large galaxy sample at $z\sim0.9$ and a selection criteria of
\oii~EW$>-$5 \AA, \citet{Schiavon2006} estimated the LTG proportion of
their sample is at most 5\%.  Though there is similarity between our
selection method and theirs, our sample size is smaller and less
complete, thus we estimate the LTG portion of our sample is 5$-$25\%,
between the predictions of \citet{Schiavon2006} and
\citet{Weiner2005}.

Besides a few bright, red LTGs in the sample, there may also be some
low power AGN. The Baldwin$-$Phillips$-$Terlevich (BPT) diagnostic
diagram \citep{Baldwin1981} suggested that \oiii~EW$<-5$ \AA, or
\ha~EW$<-5$ \AA, which is used as one of our sample selection
criteria, cannot effectively eliminate the AGN contributions.

\subsection{Composite Spectra and Index Measurements}

We retrieved the one dimensional spectra from the DEEP2 Data Release 4
website\footnote{http://deep.ps.uci.edu/DR4/spectra.html}. The reduced
two dimensional spectra obtained from DEIMOS are flat-corrected and
wavelength-calibrated. The pipeline also takes care of the sky
subtraction and cosmic ray rejection. One dimensional spectra are
extracted from each of the slitlets using an optimal extraction method
\citep{Horne1986} that assumes a constant Gaussian profile at all
wavelengths, which implies the spectral extraction region is the whole
visible galaxy. Flux calibration is not attempted along the process,
and the flux unit is DEIMOS counts per hour (e$^-$/hour).

\begin{figure}
\centering
\includegraphics [scale=0.4]{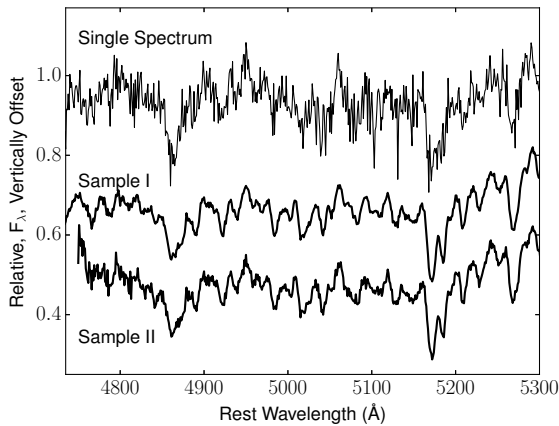} 
\caption[Spectra before and after stacking.]{A swath of DEEP2 spectra
  before and after stacking to illustrate S/N improvement. Sample I
  stacks 275 spectra, and Sample II stacks 72 spectra to achieve
  higher S/N at cost of the loss of individual galaxy
  identities.}\label{fig:stack}
\end{figure}

To stack the spectra we used pipeline programs developed by the DEEP2
team\citep{Cooper2012}. We modified the ``coadd\_spectra'' program to
meet our (mild) need for flux correction.  In measuring spectral
indices, scalings and even linear corrections do not affect the
result. However, if the response curve correction is more complicated
than linear, there is a mild effect, and so we included the flux
correction.  We divided each spectrum by the throughtput of DEIMOS in
spectroscopic mode for the gold 1200-line/mm
grating\footnote{http://www.ucolick.org/$\sim$ripisc/results.html}. Next,
each individual spectrum is shifted to the rest-frame and normalized
by dividing the median spectrum. The composite spectra are achieved by
coadding the normalized spectra, where the inverse variance of each
pixel is used as weight (see Figure \ref{fig:stack}).

We estimated the velocity dispersions ($\sigma$) of the composite
spectra in two ways; the \citet{1976ApJ...204..668F} $\sigma-L$
scaling relation and cross-correlation.  For luminosity, we
K-corrected the R band magnitude to B band magnitude using the code of
\citet{Blanton2007}. The velocity dispersion was calculated by
adopting the Faber-Jackson relation presented in
\citet{Whitmore1981}. The average $\sigma$ of our 302 red galaxy pool
is about $235 \pm 40$ km s$^{-1}$.
For the latter estimate, we cross-correlated the composite spectra
with model stellar spectrum templates, with the width of the
cross-correlation function fitted with a Gaussian and treated as in
\citet{1979AJ.....84.1511T}. Composite spectra of Sample
I and Sample II show $\sigma\sim$241 km s$^{-1}$, and $\sim$231 km
s$^{-1}$, respectively.  The velocity dispersions determined by these
two methods agree.

We broadened the spectra to 300 km s$^{-1}$
($\sigma_{broaden}^2=\sigma_{300}^2-\sigma_{sample}^2$) , and measured
spectral indices and associated errors propagated from the errors on
each flux pint. In terms of studying indices rather than the spectra
directly, we chose indices because more theoretical predictions are
available, systematic fluxing issues are minimized, and analysis is
more secure and direct. Our index table consists of Lick-style indices
presented in \citet{Worthey1994b,Trager1998, Serven2005}, and Table
\ref{tab:index}.  The measured indices and errors are shown as red
(Sample I) and light grey (Sample II) filled circles with error bars
in Figure \ref{fig:comp}. The errors plotted are pixel by pixel
measurement errors propagated through to the indices (and do not
include a contribution from velocity dispersion uncertainty that
modestly affects narrower indices).

The H$\beta$ index was corrected for emission via the methods in
\citet{Serven2010}, that is, via measurement of an H$\alpha$ index,
then using Mg $b$ to estimate a continuum slope correction. The
corrections were substantial since the stacked H$\alpha$ indices were
generally in emission, from 0.5 \AA\ for the low-luminosity subsample
up to 1.1 \AA\ for high luminosity.

We also subdivide the sample by luminosity at $m_R = 20.5$
mag. Cross-correlation indicates a 15\% smaller velocity dispersion in
the faint subsample and a 5\% larger velocity dispersion in the bright
subsample compared to the average. These differences were propagated
through the smoothing and index measurement procudures.

\section{Models and Local Observables for Comparison}
\label{sect:observe}

Stellar population model information and additional observational material is
needed for a fair comparison of IMF indicators.

\textbf{Worthey models:} Models \citep{Worthey1994a,Trager1998} that
start with empirical stellar libraries, then use synthetic spectra
\citep{Lee2009} to gauge the effects of detailed chemical composition
were employed, with a few ongoing improvements.

For this version, the isochrones of \citet{Bertelli2008,Bertelli2009}
with the thermally-pulsing asymptotic giant branch (TP-AGB) treatment
described in \citet{Marigo2008} were employed. This set of isochrones
has a low mass limit of 0.15 M$_\odot$. In philosophy similar to
\citet{Poole2010}, indices were measured from four stellar spectral
libraries
\citep{Valdes2004,Worthey2014x,SanchezBlazquez2006miles,Rayner2009},
all transformed to a common 200 km s$^{-1}$ spectral
resolution. Multivariate polynomials were fit over five overlapping
temperature zones as a function of $\theta_{eff}$ = 5040/T$_{eff}$,
log g, and [Fe/H], then smoothed and summarized in a lookup table. To
compute the integrated properties of the final model, the isochrones
plus an IMF give the numbers of stars in each bin of the
isochrone. The stellar index was found (and any optional chemical
element tweaks imposed), then decomposed into ``index'' and
``continuum'' fluxes, which were separately added, then, after
summation, re-formed into an index representing the integrated
value. To transform from 200 km s$^{-1}$ to 300 km s$^{-1}$ small
additive corrections for each index were estimated by Gaussian-broadening high
resolution synthetic composite spectra.

For this work, we updated the models with additional IMF options. We
calculate SSP models with the \citet{Kroupa2001} IMF to represent LTGs and two power law IMF slopes: $\alpha=$2.35,
3.0. For each IMF slope, our SSP models are given at the ages of 5, 10
Gyr, with [M/H] $=\{-0.33$, 0, 0.37$\}$.

\textbf{CvD12:} We also employ the Stellar Population Synthesis (SPS)
presented in \citet{Conroy2012a} (CvD12).  CvD12 models scaled-solar
old stellar populations ($>$3 Gyr) with empirical spectral libraries:
MILES \citep{SanchezBlazquez2006miles} and IRTF \citep{Cushing2005,
  Rayner2009}. Within the limitation of near-solar abundance, they
also probe individual abundance variations and $\alpha$ element
enhancements with synthetic spectral libraries as well as provide IMF
variation options.  For our illustrations, we adopt the spectra
presented in CvD12 from http://scholar.harvard.edu/cconroy/sps-models.
These spectra are the SPS model output at four ages (3, 5, 7, and 9
Gyr) with four different types of IMFs ($\alpha=3.5$, $\alpha=3.0$,
Salpeter, and Chabrier).  We measured indices after we broadened the
spectra to 300 km s$^{-1}$.

\textbf{Local galaxies:} As stellar population models still suffer
from uncertainties \citep{Charlot1996,Conroy2009,Tang2014}, local
galaxies are useful for empirical comparison. Since our sample
galaxies consists of a majority of ETGs and a few LTGs,
templates of both galaxy types are desired. Note that spectra from all
sources were broadened to 300 km s$^{-1}$ for purposes of comparison.
\begin{enumerate}
\item{ We were kindly provided with the updated mean spectra of
  non-star forming, non-LINER galaxies by G. Graves.  Galaxies from
  SDSS DR7 were selected with the following criterion: (1)
  $0.025<z<0.06$; (2) No detectable H$\alpha$ or \oii~3727 emission
  \citep{Peek2010}; (3) Median S/N$>$5 per \AA. Then, the galaxies
  were divided into six bins in $\log\sigma$: 1.86$-$2.00,
  2.00$-$2.09, 2.09$-$2.18, 2.18$-$2.27, 2.27$-$2.36, 2.36$-$2.50; (4)
  further culling if the galaxies were far from the fundamental plane.
  See \cite{Conroy2014} for a more detailed description of the sample,
  where the last velocity dispersion bin of our sample are sub-divided
  into two. Finally, H$\beta$ was corrected via \citet{Serven2010}. }

\item{ \citet{Dobos2012} classified the galaxies of SDSS DR7 by both
  colour and nuclear activity. We select the two most relevant
  subsamples: 1) The passive galaxies (with no H$\alpha$ detection)
  split by color into red, green, and blue sub-samples, and 2) a red
  sample, sub-divided into five smaller samples based on nuclear
  activity from low to high: (Their RED 1, RED 2, RED 3, RED 4, and
  RED 5 samples). In order to avoid Malmquist bias, each sub-sample is
  constrained by redshift and absolute magnitude to ensure
  volume-limited sampling. The detailed redshift and absolute
  magnitude ranges can be found in their Table 2.  For example, the
  passive samples are selected inside $0.03<z<0.14$,
  $-20.5<M_r<-21.5$. Note that this sample is more distant than the
  Graves sample. Applying the Faber-Jackson law to this sample, we
  infer velocity dispersions around 142 km s$^{-1}$ with narrow
  variance, so we adopt this value for purposes of correcting to 300
  km s$^{-1}$ for all the subsamples. H$\beta$ corrections were
  applied similar to the DEEP2 sample. Note that the Dobos passive
  sample's selection criteria are the most similar to our DEEP2
  galaxies. This point becomes important when we discuss IMF
  evolution.}

\item{ We also retrieved the Sb galaxy template from
  \citet{Kinney1996}. This optical template is a combination of two Sb
  galaxies, NGC 210 and NGC 7083, whose spectra were obtained at the
  CTIO 1 m telescope with the two-dimensional Frutti detector. The
  CTIO spectra covers 3200$-$10000 \AA~with a resolution of 8 \AA ,
  which was then smoothed to 300 km s$^{-1}$ with a smoothing kernel
  that varies with wavelength. H$\beta$ corrections were applied similar to the DEEP2 sample. }
\end{enumerate}

\section{Results}
\label{sect:comp}

\begin{figure*}
\centering
\includegraphics [width=\textwidth]{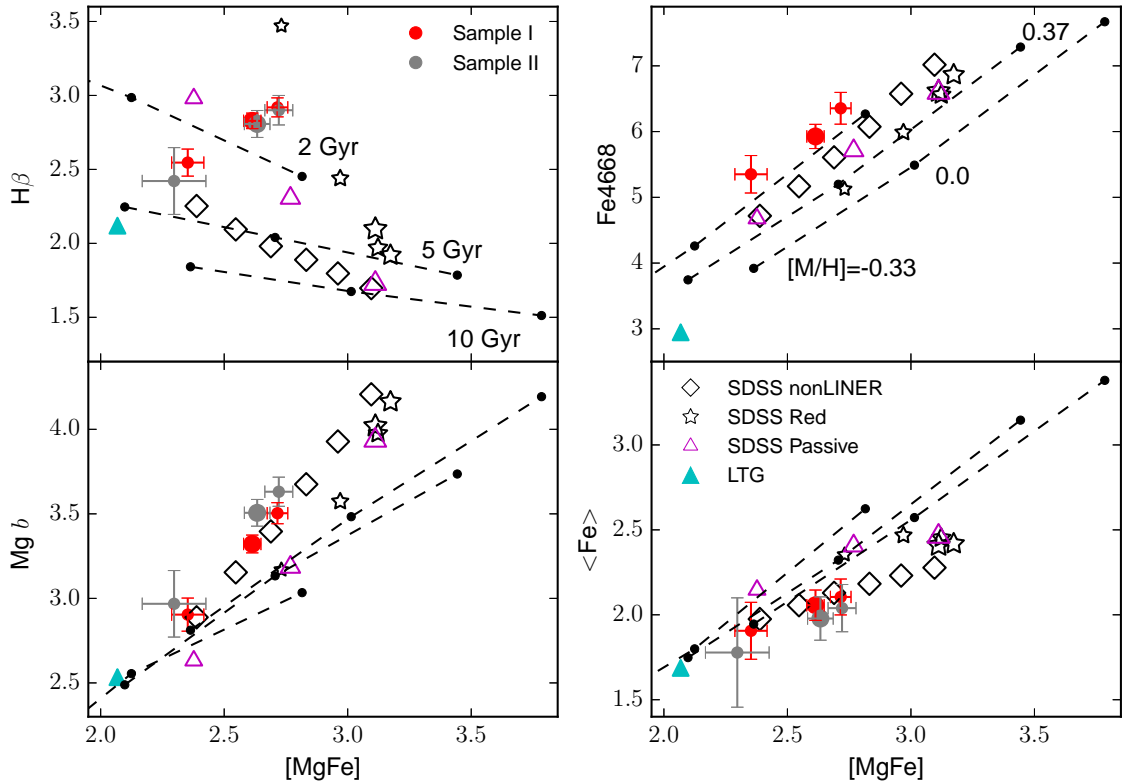} 
\caption[Comparing observables and models]{Index planes that are not
  sensitive to the IMF. They show expected age and abundance trends
  for the most part. Our SSP models are given at age$=$ 2, 5, and 10 Gyr
  with [M/H]$=-0.33$, 0, and 0.37 (dashed tracks). The observables are
  the Graves non-LINER averages binned by velocity dispersion (open
  diamonds), Dobos Passive split into three color bins (open magenta
  triangles), Dobos RED split by nuclear activity (black five-pointed
  stars; the smallest symbol has the least nuclear activity, the
  largest the most), Kinney LTG (cyan filled triangle), \oiii selected Sample I (red
  filled circles), and H$\alpha$ selected Sample II (grey filled circles). Each Sample
  I/II grand average is indicated by a larger symbol, and the smaller
  symbols indicate a split in the sample at magnitude
  $m_r=20.5$.}\label{fig:precomp}
\end{figure*}

\begin{figure*}
\centering
\includegraphics [width=\textwidth]{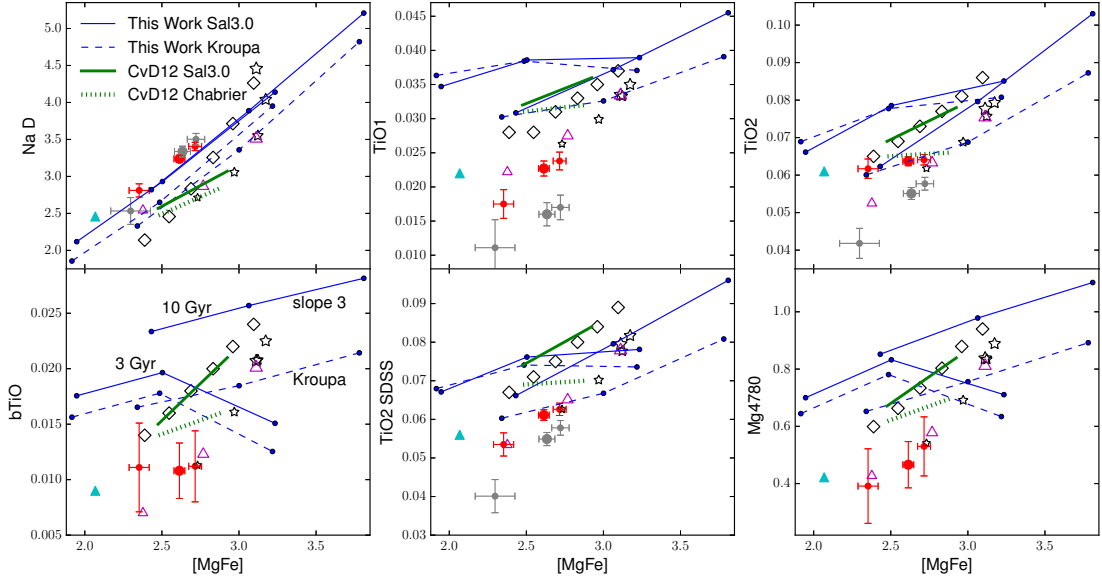} 
\caption[Comparing observables and models]{Index planes that are
  sensitive to the IMF. Most symbols are as in
  Fig. \ref{fig:precomp}. The selection of models is altered,
  however. Our SSP models are shown for ages 3 and 10 Gyr, and for a
  bottom-heavy $\alpha = 3.0$ power-law IMF (solid blue lines) and a
  Kroupa IMF (dashed blue lines) that is similar to the Chabrier
  IMF. In addition, CvD12 models are given with lines connecting ages
  3 and 9 Gyr at solar metallicity. An $\alpha=3.0$ power-law IMF
  (solid green line) and a Chabrier IMF (dotted green line) give the
  IMF sensitivity. If both sets of models agreed perfectly, the CvD12
  models would nearly connect the middle two dots on the
  metallicity-sensitive model isochrones.}\label{fig:comp}
\end{figure*}

Figures \ref{fig:precomp} and \ref{fig:comp} intercompare index values
from the models and the observations. The [MgFe]\footnote{
  [MgFe] $\equiv \sqrt{{\rm Mg}\ b *( {\rm Fe5270+Fe5335})/2}$, from
  \citet{Gonzalez1993}} is chosen as the $x$-axis variable since it tracks mostly [M/H] rather than [$\alpha$/H] or [Fe/H], though it retains considerable sensitivity to population age.

Examining Fig. \ref{fig:precomp}, the top-left panel shows
age-sensitive H$\beta$. The model grid show two isochrons at 2, 5, and
10 Gyr with dots at [M/H] $= -0.33$, 0.0, and 0.37. Grid
extrapolations are approximately linear. Variable IMF models are not
shown in Fig. \ref{fig:precomp}, and the IMF is a power law with low mass cutoff of 0.15 M$_\odot$. The DEEP2 galaxies are marked with
circles and error bars. The two sample grand averages, marked with larger
circles, are flanked by smaller circles that represent a split of the
sample at $R=20.5$. This is approximately half by number, although the fainter
half has larger errors. The Graves passive galaxies (open
black diamonds representing a series of bins in velocity dispersion),
Dobos passive galaxies (open triangles representing three color bins),
and Dobos RED galaxies (open stars representing AGN activity
increasing from small symbols to large symbols) are also plotted.

In terms of velocity dispersions, the Dobos galaxies locate about the
3rd (from the left) Graves bin, while the DEEP2 galaxies locate
between the 5th and 6th. The DEEP2 galaxies should be younger by $\sim
4.3$ Gyr due to lookback time, all else being equal. The H$\beta$
emission corrections for the active galaxies are more uncertain
because the corrections \citep{Serven2010} were built for star
formation scenarios, not AGN.

The trends apparent in the H$\beta$ diagram are that more
massive/redder galaxies appear older and more metal rich relative to
the model grid, a trend seen since the invention of this diagnostic
diagram \citep{1992IAUS..149..255F}. There appears to be only one
contradiction apparent in the H$\beta$ diagram, which is that the
Dobos sequences might be expected to cross the Graves sequence at the
third diamond (matching velocity dispersions), not converge at the
6th, as observed. However, that is only a contradiction if the samples
are expected to be equivalent. If we posit that sample selection can
have an influence, then what we see are diagnostics of the sample
selection method. Points to note in the H$\beta$ diagram:

\begin{itemize}
  \item{ETGs form a narrow sequence in [M/H] but range over a large range of average age.}
  \item{The LTG template had a large nebular emission correction, but
    within that uncertainty appears to be of intermediate age but a
    few tenths more metal poor than the ETGs. }
  \item{The three samples (Graves nonLINER, Dobos Passive, Dobos Red) tilt progressively such that the bluer Graves galaxies are more metal poor, the bluer Dobos Passive galaxies are on an isometallicity track, and the bluer Dobos Red galaxies (the least AGN active) are more metal rich.}
  \item{The DEEP2 samples lie slightly more metal rich from the convergent red end of the other samples, but much younger. The youthening effect is much stronger than passive evolution. }
\end{itemize}

Moving to the other three panels of Fig. \ref{fig:precomp}, these are
diagrams typically used to track abundance ratio changes because they
are age-metallicity degenerate \citep{1998PASP..110..888W}. The usual
conclusion is that smaller ETGs resemble LTGs in their abundance
ratios, and that these ratios are scaled-solar. From there, the
massive ETGs show enhanced Mg and other light elements, while Fe-peak
elements are depressed relative to the average. These trends are
confirmed here. It is statistically significant that, except for the
reddest few, the Dobos samples are a bit more Fe-rich, Mg-poor, also
lie low in Fe4668, which is mostly due to carbon. This must be sample
selection. Mentally allowing for the age difference, which weakens all
the metallic feature indices, the DEEP2 sample shares the typical abundance
pattern of massive ETGs.

Fig. \ref{fig:comp} shows IMF-sensitive indices. In both sets of
models, a young population age decreases sensitivy to IMF. The two
sets of models agree qualitatively, though the present model set is
always shifted a bit more strong-lined than CvD12. In all the various
indices pictured, shallow IMFs imply weaker feature
strengths\footnote{ Exceptions to this general behaviour is found in
  the youngest and most metal-poor populations for TiO1, TiO2, and
  TiO2-SDSS indices. This is because the IMF effects in young,
  metal-poor populations and old, metal-rich populations are dominated
  by stars from different stellar phases. Readers are referred to
  \citet{Tang2015} for more details.}. Looking at the diagrams
empirically, except for the least-diagnostic Na D panel, the Graves
sequence is stronger-lined in all IMF diagnostics. The empirical
points seem to split into two families, with the Graves sample
defining one, and all the others, including the LTG template, Dobos
samples and DEEP2 ETGs, defining the other.  That the Dobos samples
and DEEP2 ETGs share the same shallow IMF indicates little cosmic
evolution in IMF over the last 4 Gyr.

For disentangling possible age effects, perhaps the best diagrams are
the TiO2 and TiO2-SDSS indices, because the models begin to develop
strong AGBs at younger ages so that the models never dip weak enough
to reach the observed DEEP2 or Dobos indices unless the IMF is
shallow. For disentangling abundance ratio effects, we look for the
various TiO diagrams to be echoed in the Mg4780 diagram. The
morphology is the same, and we conclude that IMF may play a role in
the index strengths.

The Na D index has a small amount of IMF sensitivity, but it has a
large amount of sensitivity to abundance ratios. It may also,
especially in the cases of the LTG template and the DEEP2 sample,
suffer from galaxy self-absorption if neutral Na is present in the
interstellar medium, since both the component of Na D are resonance
lines. The Graves and Dobos sequences tilt such that
more massive ETGs have a steeper IMF than the lightweight ones.

\section{Discussion, Summary, and Conclusion}
\label{sect:disc}

Random uncertainties are shown in Figs. \ref{fig:precomp} and
\ref{fig:comp} except that the random uncertainties are smaller than
the point symbols for all the SDSS averages and could not be estimated
with certainty for the LTG template, but are plausibly of order 0.1
\AA\ or 0.005 mag.

Systematic uncertainties are a bigger worry. Velocity dispersion
corrections are important for H$\beta$, Mg $b$, $<$Fe$>$, Na D, and
Mg4780, but a much lesser concern for Fe4668 and the four TiO
indices. Corrections were applied to all indices, however, so the
primary uncertainty is the difficulty of knowing the appropriate
velocity dispersion to assign to each averaged bin, weighted by how
close the galaxy bin is to the target 300 km s$^{-1}$, because the
closer the galaxy is to that target, the smaller the correction. We
judge that the only data points that might suffer a significantly
skewed result are the low-luminosity subsamples from DEEP2. There are
additional systematics for the H$\beta$ diagram due to emission corrections, as
discussed above, and a small worry for extra Na D self-absorption, a few
tenths of an Angstrom at most
\citep{1983MNRAS.204..317B,1986A&A...166...83B}. Milky Way Na
absorption is not an issue due to the redshifts of the galaxies. The most
serious uncontrolled systematic effect is likely to be spectrophotometric
integrity. If spectral response curvatures the same order as the index
widths occur, and are not averaged away, they will produce spurious
unastrophysical drifts in the indices. The fluxing procedures in DEEP2
and the averaging in SDSS samples will minimize this. The indices most
prone to fluxing errors are TiO1, TiO2, and TiO2-SDSS due to their long
wavelength spans. Indeed, it is in these indices that Sample I and
Sample II diverge strongly, and that is probably no coincidence.

Systematic effects in the models are likely to be more severe. Fluxing
should not be a problem, but velocity dispersion corrections must be
made. Furthermore, besides simplified, parameterized IMFs, the models
are based on stellar evolutionary models which, while admirable in
many ways, are also prone to uncontrolled drifts in stellar
temperatures, luminosities, and lifetimes \citep{Charlot1996}. We
recommend eyeing the models only in a differential sense, looking for
the vectors of age, [M/H], and IMF in the various index-index diagrams.

A few addtional considerations deserve a few words. \citet{Tang2015}
studied two effects that might entangle with the IMF slope
determination, namely the IMF Low Mass Cut-Off (LMCO), and AGB
contribution effects. The degeneracies between slope, LMCO, and AGB
star contributions are real, but testing indicates they are too subtle
to affect the appearance of Fig. \ref{fig:comp}.

In ETGs, [Ti/Fe] $\approx$ 0
\citep{Johansson2012,Conroy2014,Worthey2014}. Could Ti be even more
underabundant in our galaxies? Additional indices Ti4553 and Ti5000
(not illustrated and not sensitive to the IMF) lie slightly lower than
solar for our sample galaxies. Thus the small values of Ti-related
indices may be partially due to the low Ti abundances. However,
sub-solar Ti abundances can only explain indices containing Ti.  The
Mg4780 index is not Ti-sensitive, yet indicates a shallower IMF even
without accounting for Mg-enhancement.  Mg4780 is defined in
\citet{Serven2005}, and shown to be IMF-sensitive in
\citet{LaBarbera2013}. Our models accomodate varying Mg and Ti
separately, but testing with abundance-altered models fails to
indicate any easy way to relieve the appearance that shallow IMFs are
required in Fig. \ref{fig:comp}.

Since the SDSS spectra are limited to the galaxy center by the fixed
fiber size (3 arcsec), the relatively nearby ($0.025<z<0.06$) Graves
stack may be biased towards the stellar populations in the galaxy
center. The Dobos samples are less affected by the aperture effect,
since the sampling galaxies are on average further away ($0.03<z<0.14$
for the passive samples). For the DEEP2 samples, the long-slit
spectroscopy and the optimal extraction method ensure that the DEEP2
spectra are free from aperture effect. According to the recent work of
the MaNGA team on stellar population gradients
\citep{Zheng2016,Goddard2016}, no or slightly negative metallicity
gradients are found in a large sample of nearby galaxies, though the
metallicity gradients may be stronger in more narrowly selected
massive ETGs \citep{VanDokkum2016}. Therefore, the Graves stack
measurements with larger $\sigma$ may be slightly affected by the
metallicity gradient. The correcting vectors should point to the
lower-left corner of the panels in Figure \ref{fig:comp}, indicated by
our SSP models (blue lines). Furthermore, the IMF gradients suggested
by \citet{MN2015,VanDokkum2016} may also affect the the Graves stack
measurements with larger $\sigma$, thus the aperture effect may
magnify the IMF-related indices.

\subsection{Summary}
\label{sect:con}

With these caveats in mind, the firm conclusions of examining
Figs. \ref{fig:precomp} and \ref{fig:comp} are

\begin{itemize}
  \item{Sample selection strongly drives every diagnostic diagram.}
  \item{The brightest, reddest galaxies are the oldest on
    average. From there, however, sample selection drives a newly seen
    trend for age. In the sample selected by ETG fundamental plane,
    bluer galaxies are more metal-poor. In a sample composed purely of
    non-AGN, non-star-forming galaxies, bluer galaxies are the same
    metallicity as red ones but much younger. In a sample with
    detectable AGN activity, the bluer galaxies are more metal rich,
    and that partially anticorrelates with AGN activity; the
    least-active galaxies are both younger and more metal rich.}
  \item{To explain all the index drifts in comparison to the models,
    IMF variations seems to be required, along with age and abundance
    variations. In all SDSS averages there is trend that the strongest
    lined galaxies appear to have a more bottom heavy IMF, while the
    weaker lined galaxies have a spiral-like bottom light IMF, in
    accord with many recent studies.}
  \item{The SDSS sample that is culled to be near the fundamental
    plane of ETGs and thus is probably the purest in terms of ETG
    fraction forms a separate sequence that is offset toward a
    bottom-heavy IMF. The other SDSS samples that likely contain more
    LTGs group with the similarly-selected DEEP2 samples and the LTG
    template galaxy itself for a sequence at a more bottom-light location.}
  \item{Split by luminosity, the low-luminosity half of the DEEP2 red
    galaxy sample is slightly more metal-poor, which is expected if
    metallicity drives the color-magnitude relation, but also older on
    average, which is an interesting new result that should be
    confirmed when more and better spectra of high redshift galaxies
    are available.}
  \item{The DEEP2 samples appear to be slightly enhanced in carbon
    (Fe4668 index) and sodium (Na D index) and quite possibly slightly
    deficient in titanium (the various TiO indices) compared to their
    zero-redshift cohorts.}
  \item{The DEEP2 samples are about a factor of two younger than would
    be inferred if they were the passively evolving precursors of the
    nearby strong-lined galaxies. That is, in Fig. \ref{fig:precomp},
    galaxies at the DEEP2 velocity dispersion have ages around 8 Gyr,
    the DEEP2 galaxies have ages less than 2 Gyr, and the lookback
    time is roughly 4 Gyr.}
  \item{Since the DEEP2 samples mesh with similarly-selected nearby
    galaxies (the Dobos samples), we do not find evidence of cosmic
    evolution in IMF over the last 4 Gyr. However, our fidelity at
    detecting IMF variations is low, and such evolution could exist at
    a modest level.}
\end{itemize}

\subsection{Future Improvements}

The spectroscopic study of IMF cosmic evolution can be improved in
several ways:
\begin{enumerate}
\item{ Studying galaxies at higher redshift would presumably
  accentuate trends seen at less extreme lookback times. Given that
  DEEP2 Galaxy Redshift Survey is designed for galaxies at $z\sim1$,
  the wavelength range we see is the near-ultraviolet.  Finding
  IMF-sensitive indices in the ultraviolet, however, is formidable, if
  not impossible, since the diagnostic cool stars emit only a small
  percentage of their light there.}

\item{ Alternatively, keeping to modest redshifts of $0.3<z<0.5$, a
  larger sample would increase the S/N ratio of the composite spectra
  and allow subdivision of samples to better characterize any IMF
  evolution. Morphological classification, fundamental plane
  subselections, and AGN information would improve the certainty of
  what sort of objects are under study. SDSS-BOSS
  \citep{2016MNRAS.tmp.1473R} could be mined, for example. }

\item{ Spectroscopic study of the IMF in the J band has recently
  become possible, and has been successful \citep{Smith2015}.  At
  z=0.4, the \nai~and \caii~features would fall into the J-band,
  therefore these classical feature lines can be used for future
  cosmic evolution study.}
\item{ Targeting individual distant galaxies at high S/N would yield
  information on galaxy to galaxy cosmic scatter in IMF parameters, as
  well as age and abundances, perhaps revealing a whole new level of
  detail about galaxy evolution. It may be feasible to observe
  a sample of bright galaxies in a cluster with a multiplexed instrument
  (e.g., VLT/KMOS or Keck/MOSFIRE) in a reasonable time. }
  
\end{enumerate}

\subsection{Conclusion}

Recent research on nearby galaxies suggests variable IMFs and
questions the universality of the IMF seen around the solar
neighbourhood. In particular, massive elliptical galaxies appear to
have a bottom-heavy IMF in comparison to low-mass elliptical or spiral
galaxies. For high-redshift galaxies, the red IMF-sensitive
indices shift out of the CCD wavelength range.  In this paper, we
apply a set of bluer IMF-sensitive indices to the studies of
intermediate-redshift ($0.3<z<0.5$) galaxies.  Red galaxies are
selected from the DEEP2 Galaxy Redshift Survey and stacked.  Spectral
indices measured from the composite spectra are compared with two sets
of models and also local galaxies.

We confirm recent work that strong-lined (red, massive) galaxies
appear to have a bottom-heavy IMF compared to weak-lined ETGs and
LTGs.  There is an affirming additional trend that the local sample
that culls out fundamental plane outliers and thus selects what we
think of as elliptical galaxies the best is offset from samples that
are selected in more inclusive ways, in the sense that the pure-E
sample is offset toward a more bottom-heavy IMF. The DEEP2 galaxies do
not appear that bottom-heavy, joining local LTGs and
similarly-selected SDSS averages. There is no evidence for evolution
in the IMF over the last 4 Gyr, at least with current data and tools.

In terms of ages and abundances for local galaxies, sample selection
drives a fascinating trend in which all of the reddest galaxies
converge at metal-rich and old, but bluer galaxies do not agree:
fundamental plane culled blue galaxies lie more metal poor and a bit
younger, zero-nebula selected blue galaxies lie at the same abundance
but much younger, and the bluest bins of red galaxies with detectable
AGN lie much younger and \textit{more} metal rich.

The DEEP2 red galaxies, if split in half by luminosity, show that the
faint half is more metal poor, and, surprisingly, older. This trend
may be driven by small number statistics. The DEEP2 red galaxies are
also quite young, less than 2 Gyr, as compared to 8 Gyr for local
galaxies at the same velocity dispersion, effectively ruling out a
passive evolution hypothesis. The DEEP2 red galaxies are
likely slightly enhanced in C and Na and slightly deficient in Ti.
 
\section{Acknowledgement}

B. T. would like to thank Y. Chen for her help on the DEEP2 spectral
reduction and J. Newman and R. Yan for useful technical
advice. We thank the referee, Russell Smith, for insightful comments.
Support to G. W. for program AR-13900 was provided by NASA
through a grant from the Space Telescope Science Institute, which is
operated by the Association of Universities for Research in Astronomy,
Inc., under NASA contract NAS 5-26555.

\bibliographystyle{mn}
\bibliography{imfdeep2}

\label{lastpage}

\end{document}